\documentclass[acmlarge]{acmart}

\usepackage{framed,multirow}

\usepackage{amssymb, amsmath}

\usepackage{latexsym}
\usepackage[T1]{fontenc}
\usepackage{siunitx}
\usepackage[]{algorithm2e}
\usepackage{subcaption}

\usepackage{url}
\usepackage{xcolor}

\usepackage{hyperref}

\usepackage{graphicx}
\graphicspath{ {images/} } 

\definecolor{newcolor}{rgb}{.8,.349,.1}
\AtBeginDocument{%
  \providecommand\BibTeX{{%
    \normalfont B\kern-0.5em{\scshape i\kern-0.25em b}\kern-0.8em\TeX}}}

\setcopyright{acmcopyright}
\copyrightyear{2020}
\acmYear{2020}

\acmJournal{TMIS}

\begin{document}

\title{CoRSAI: A System for Robust Interpretation of CT Scans of COVID-19 Patients Using Deep Learning}

\author{Manvel Avetisian}
\authornote{Corresponding author: avetisian.m.s@sberbank.ru}
\authornote{Authors listed in alphabetical order.}
\email{avetisian.m.s@sberbank.ru}
\affiliation{
  \institution{Sberbank AI Laboratory}
  \streetaddress{Oruzheynyy Pereulok, 41}
  \city{Moscow}
}


\author{Ilya Burenko}
\email{burenko.i.m@sberbank.ru}
\affiliation{
  \institution{Sberbank AI Laboratory}
}

\author{Konstantin Egorov}
\email{egorov.k.ser@sberbank.ru}
\affiliation{
  \institution{Sberbank AI Laboratory}
}

\author{Vladimir Kokh}
\email{kokh.v.n@sberbank.ru}
\affiliation{
  \institution{Sberbank AI Laboratory}
}

\author{Aleksandr Nesterov}
\email{AINesterov@sberbank.ru}
\affiliation{
  \institution{Sberbank AI Laboratory}
}

\author{Aleksandr Nikolaev}
\email{a.e.nikolaev@yandex.ru}
\affiliation{}

\author{Alexander Ponomarchuk}
\email{ponomarchuk.a.v@sberbank.ru}
\affiliation{
  \institution{Sberbank AI Laboratory}
}

\author{Elena Sokolova}
\email{Sokolova.El.Vladimirov@sberbank.ru}
\affiliation{
  \institution{Sberbank AI Laboratory}
}

\author{Alex Tuzhilin}
\email{atuzhili@stern.nyu.edu}
\affiliation{
  \institution{Sberbank AI Laboratory}
}
\affiliation{
    \institution{New York University}
    \streetaddress{New York, NY 10003, United States}
    \city{New York}
}

\author{Dmitry Umerenkov}
\email{D.Umerenkov@gmail.com}
\affiliation{
  \institution{Sberbank AI Laboratory}
}


\begin{abstract}
    Analysis of chest CT scans can be used in detecting parts of lungs that are affected by infectious diseases such as COVID-19. Determining the volume of lungs affected by lesions is essential for formulating treatment recommendations and prioritizing patients by severity of the disease. In this paper we adopted an approach based on using an ensemble of deep convolutional neural networks for segmentation of slices of lung CT scans. Using our models we are able to segment the lesions, evaluate patients dynamics, estimate relative volume of lungs affected by lesions and evaluate the lung damage stage. Our models were trained on data from different medical centers. We compared predictions of our models with those of six experienced radiologists and our segmentation model outperformed most of them. On the task of classification of disease severity, our model outperformed all the radiologists.
\end{abstract}



\maketitle

\section{Introduction}

Coronavirus (COVID-19) has spread widely around the world since the beginning of 2020, and an extensive effort to combat the pandemic was launched that year. As a result of this effort, there have been several diagnostic tests developed in the medical community to detect the COVID cases.
One of the most prominent methods to confirm a COVID-19 infection is by conducting the reverse transcriptional polymerase chain reaction (RT-PCR) test, which 
has a lower sensitivity of 65-95\%. 
Although useful and popular, the RT-PCR  test has the problems of producing negative results even if the patient is infected, and also the problem of waiting for the test results.
Therefore, in some countries a chest computed tomography (CT) scan is widely used in clinical practice to detect typical changes in the pulmonary parenchyma associated with COVID-19 \citep{ref_article1, ref_article4, ref_article5, ref_article6} as a complement to the RT-PCR test, especially since CT is effective for early detection and diagnosis of COVID-19 \citep{ref_article7, ref_article8} and since the results of CT scans can be analyzed immediately \citep{ref_article2, ref_article3}. Multifocal ground-glass opacifications (GGO) is the most common finding of the CT scan, usually localized peripherally in both lungs, while  single ground-glass lesion can be common at an early stage of the disease \citep{ref_article9}. Clinical manifestations of COVID-19 pneumonia and their severity correlate with the volume of lung damage, which can be assessed using visual or quantitative scale.

Although it is easy to assess the severity of lung damage using a visual scale, this is a subjective assessment which can vary substantially among radiologists \citep{ref_article7}. Therefore, there exists a more objective classification of lung damage widely used in some countries, including Russia, consisting of the following five stages (referred in the paper as CT-classes): CT-0: absence of damage; CT-1: pulmonary parenchymal involvement (PPI) being $\leq$ 25\%; CT-2: PPI being in the range of 25-50\%; CT-3: PPI in the range of 50-75\%; and CT-4: PPI $\geq$ 75\% \citep{ref_article6}.
In the context of the current COVID-19 pandemic, radiologists in specialized departments need to process a very large number of CT images of subjects with suspected COVID-19, sometimes up to several hundred patients per day, which puts incredible burden on them and also delays the COVID-19 detection event. Therefore, an automated system 
that can accurately detect the presence of COVID-19 and calculate the pathology of lung volume will significantly reduce the burden on the radiologist, help objectively assess the severity of the disease, make it possible to prioritize the radiologist work schedule, and provide better insights into the follow-up studies to assess the dynamics of the disease.



In this paper, we present the CoRSAI system\footnotetext{CoRSAI stands for \textit{R}u\textit{S}sian \textit{CO}ronovirus \textit{AI}-based detection system} that takes CT scans of COVID-19 patients 
and does the image classification and segmentation tasks using Deep Learning based methods to find the affected areas, to determine severity of the disease, and to track disease progression. The proposed system uses a novel ensemble of previously developed DL-based models that was architected specifically with the goal of detecting lung damages caused by COVID-19.

To test our system, we compared its performance with two existing DL-based baselines on two open datasets. As a result, our system outperformed these baselines.
In addition, we also conducted a study in which we compared CoRSAI's performance with that of six radiologists having at least three years of practical experience across the following typical tasks: 
\begin{itemize}
    \item{\it Segmentation:} detection of the affected areas of the lungs
    \item{\it Patient's dynamics:} detection of positive response to the therapy or disease progression
    \item{\it Lesion share estimation:} assessment of the lung damage share (ratio of lesion volume to lung volume)
    \item{\it Classification:} identification of lung damage stage according to the CT-class 
\end{itemize}    

We performed these four experiments using 58 CT scans on 49 patients at a large Russian hospital and used the services of 6 experienced radiologists.  

The results of this study show that our system outperformed the experienced radiologists for the segmentation and classification tasks on average. In all the cases, our system correctly determined the patients dynamics. The results of the lesion share estimation are not directly usable due to a high degree of radiologists subjectivity on this task. Correcting for this subjectivity bias, allows our system to outperform all six radiologists on the classification task, three of them with statistical significance. 


These results imply that our system can be used as a second opinion tool that would help radiologists to deal with the coronavirus pandemic. In fact, our system has been favorably received by the medical community in Russia and has been successfully deployed in several hospitals in the country.

In this paper, we make the following contributions:
\begin{itemize}
    \item First, we propose an ensemble method specifically designed for the COVID detection problem for the CT scan data that we implemented as a part of the CoRSAI system;
    \item Second, we empirically compare CoRSAI with two existing baselines and demonstrate that our method outperforms these baselines on the public and on our proprietary CT scan data;
    \item Third, we conducted a study in which the CoRSAI system outperformed six experienced radiologists across various COVID detection tasks. 
\end{itemize}

We give an overview of existing approaches to classification and segmentation of CT scans and chest X-ray studies in section \ref{sec:related_work}; in section \ref{sec:data} we give detailed description of datasets that we used for classification and segmentation tasks as well as for experiments with doctors; in section \ref{sec:methods} we describe models that we utilized, how we preprocess data and how we combine individual models into ensemble; the next section \ref{sec:experiments} is devoted to experiments that we conducted and results we obtained; we give a conclusion and some final thoughts in section \ref{sec:conclusion}.


\section{Related Work}
\label{sec:related_work}

    Using Convolutional Neural Networks (CNNs) is a common practice for the task of image segmentation. Since its appearance in 2015, the U-Net architecture \citep{u_net} and its modifications have been widely used for the medical segmentation tasks during the analysis of X-rays, CT scans, MRIs and ultrasound signals for detecting pneumonia \citep{chexnet}, breast cancer \citep{wu_breast_cancer}, stroke \citep{manvel2019radiologist}, liver tumor segmentation \citep{liver_segmentation}, prostate cancer \citep{v_net_dice_loss} and many other medical problems \citep{segmentation_survey}. 
    
    Furthermore, there is a large body of work on applying CNNs to the task of nodule detection in the chest computed tomography images  \citep{hua2015computer}, segmentation of the interstitial lung disease \citep{anthimopoulos2016lung}, chest organ segmentation \citep{charbonnier2017improving, skourt2018lung} and other related tasks \citep{lee2019deep}.
    
    There is a large body of recent work dedicated to the task of detecting COVID-19 lesions in lungs based on X-rays studies and CT scans. In particular, \citep{cell_paper} and \citep{dual_sampling} focus on differentiating coronavirus-induced pneumonia from other pneumonia types and healthy controls. Both papers describe the experiments conducted on large samples of cases and produce comparable results with high levels of differentiation between these two types of pneumonias. In \citep{weakly_supervised_framework}, a model was developed that classified whether a CT scan contains COVID-19 lesions or not, achieving ROC AUC of 0.959 based on the CT-level annotations. In \citep{infnet}, the authors describe a supervised and a semi-supervised approaches to segmentation of lesion and lungs. In \citep{JCS} a joint classification and segmentation model is built in order to achieve higher quality by extending expensive segmentation dataset with classification dataset. In \citep{MomentumContrastive} the same problem was solved by using contrastive learning to train a neural network that can later be adopted for classification task. Moreover, several publications, such as \citep{light_network, light_network2}, show that detecting coronavirus-induced lesions can be done using lightweight networks with a small number of parameters.
    
    Ensemble learning is an old and well-studied subfield of the machine and deep learning (see for example \citep{ensemblelearning1}, \citep{ensemblelearning2}, \citep{ensemblelearning3}). Several works used ensembling to improve performance of individual models. To name few in \citep{ensemblingxray1} ensembling was used to improve performance of several classification tasks using chest X-ray scans. In \citep{dual_sampling}, the authors utilize the ensemble-based approach to distinguish between the COVID-19 and the commonly acquired pneumonia on the CT scan images. In \citep{ensemblect1} and \citep{ensemblect2} authors used ensembling for the classification of CT slices. The former work uses 2-stage transfer learning, where on the first stage weights of the convolutional part of networks were frozen and only classification head were trained; on the second stage the whole pipeline was finetuned to achieve high classification score while the later utilized relative majority voting to produce a classification result. There is no published work that utilizes ensemble learning of deep convolutional models both for the classification and segmentation tasks on CT studies of lungs affected by COVID.
    
In this paper, we build on all this previous work of analyzing CT scan images by developing ensembles of the previously proposed neural networks that are specifically designed for the COVID-19 related problems. Following terminology of \citep{covidnature} we conducted experiments with internal and external validation i.e. experiments where the test data was either from the same distribution as train data or from the different distribution. Furthermore, we describe our clinical study involving several experienced radiologists on whom we test the quality of our ensemble-based model by comparing its performance with the performance of these radiologists across four coronavirus related tasks.


\section{Data}
\label{sec:data}

In our work we utilize four different anonymized CT chest datasets that we use for training, validation, testing, lung segmentation and experimentation purposes. We describe these datasets in the rest of this section.

\begin{figure}
\includegraphics[width=4.2cm]{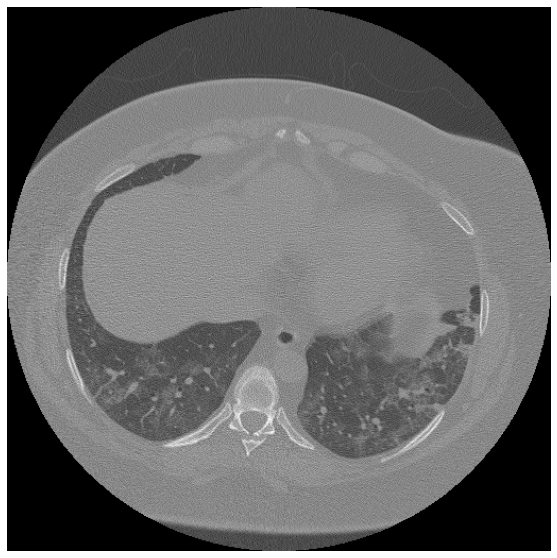}
\includegraphics[width=4.2cm]{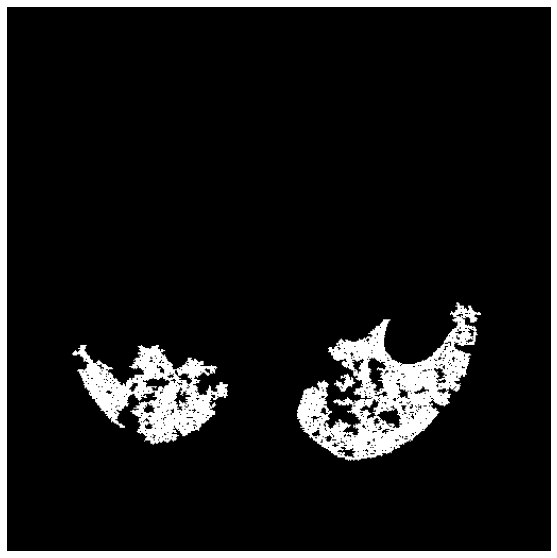}
\caption{An example of training set item: a CT scan and an annotation by radiologist}
\label{img:annotation_example}
\end{figure}

\emph{Training and validation}. This dataset consists of 68 unique anonymized CT scans with slice thickness from 0.5 to 2.5 mm collected from several hospitals and performed for the patients having COVID-19 diagnosis. It contains a collection of 18,383 original two-dimensional slices, and 9,030 segmented two-dimensional slices with lesions that we used for the training and validation purposes (see table \ref{segm_data_summary}. 
Based on the radiologist reports provided with the CT scans, we selected CT scans which have  increased lung opacity levels because of the ground-glass and consolidation findings (i.e., CT-class distribution being CT-1 (44.6\%), CT-2 (35.4\%), CT-3 (15.4\%), CT-4 (4.6\%). The CT scan series with lung window using SeriesDescription DICOM tag like «Lung» level were selected for our research. Each of these series were segmented by the radiologists in a semi-automated fashion using medical image viewer software for the segmentation purpose as follows.
First, they used the grow region feature with the lower threshold set to $-640$ and the upper threshold set to $-240$. Second, radiologists fixed the results of automated segmentation by manually making corrections to the masks on each slice for the CT scans series by using the brush/erase feature. An example of radiologist's annotation is depicted on figure \ref{img:annotation_example}.

\emph{Lung Segmentation}. To build the model of the left and the right lung segmentation, we used a subset of the LIDC/IDRI database \citep{lidc-idri-ds} from the Luna16 challenge \citep{luna16}.  This dataset contains 888 chest CT scans consisting of 227,301 normal two-dimensional slices and 194,805 segmented two-dimensional slices with the thickness levels ranging from 0.5 to 2.5 mm.

\emph{Testing}. For the model testing, we used a subset of MosMedData dataset \citep{ref_article6} containing 50 anonymized CT scans that have been annotated by the radiology experts from the Research and Practical Clinical Center for Diagnostics and Telemedicine Technologies of the Moscow Health Care Department. This testing dataset contained a collection of 2049 original and 785 segmented two-dimensional slices across 50 anonymized chest CT scans with confirmed diagnosis. See summary in table \ref{segm_data_summary}.

\begin{table}[!t]
\caption{Summary for segmentation datasets}
\label{segm_data_summary}
\centering
\begin{tabular}{|c|c|c|}
\hline
Dataset&Segmanted Slices& Total\\
\hline
Train and Validation&9030&18383\\
\hline
Test&785&2049\\
\hline
\end{tabular}
\end{table}

\emph{Experimentation}.
Finally, we prepared an additional dataset for the experimentation purposes in order to do the final comparison of the performance results of our model with that of the radiologists. 
This experimentation dataset consist of 58 anonymized chest CT scans of 49 patients. 
Since we used this dataset in four different experiments, radiologists applied different labeling methods for this dataset across these four cases.
In particular, these four types of labels are designed for the tasks of 
segmentation, dynamic classification, CT-classification and lesion share of the chest CT scans that were described in the Introduction and will further be explained in Section \ref{sec:experiments}. 
Furthermore, Table~\ref{exp_dataset} summarizes the specifics of the Experimentation dataset that has been partitioned into three subsets of sizes 20, 18 and 20 corresponding to different experiments presented in Section \ref{sec:experiments}. 

\begin{table}[!t]
\centering
\caption{Experimental subsets with labeling types.}
\label{exp_dataset}
            \centering
            \begin{tabular}{|p{1cm}|p{1cm}|p{1cm}|p{3cm}|}
            \hline
            \textbf{\#}&\textbf{Subset name}&\textbf{ \#CT scans} & \textbf{Types of label}  \\
            \hline
            First & F(20) & 20 & Segmentation\\
             & & & CT-classification\\
            \hline
             & & & CT-classification\\
            Second & S(18) & 18 & Dynamic-classification\\
             & & & Lesion share,\% \\
            \hline
            Third & T(20) & 20 & CT-classification\\
             & & & Lesions share,\% \\
            \hline
            \end{tabular}
        \end{table}



\section{Methods}
\label{sec:methods}

In this section we describe how we took the existing segmentation and classification models previously described in the literature and combined them in a unique fashion into our CoRSAI system using ensemble methods for the pneumonia-covid detection problem.
        \begin{table*}[!t]
            \centering
            \label{segm_perf_table}
            \caption{Mean DSC and standard deviation for segmentation models on test dataset}
            \centering
            \begin{tabular}{|c|c|c|c|c|c|c|}
            \hline
            &DPN& DPN-3D&FPN&ResNet-21&RN-21 + RN-18&Final\\
            \hline
            Individual model &
            $0.565 \pm 0.024$ & N/A &
            $0.572 \pm 0.016$ &
            $0.508 \pm 0.024$ &
            N/A &
            N/A \\
            \hline
            Ensemble &
            0.603$^a$ &
            0.613$^b$ &
            0.595$^a$ &
            0.601$^c$ &
            0.620$^c$ &
            0.643$^d$ \\
            \hline
          \multicolumn{6}{l}{$^a$ Ensemble on 6 folds.}\\
          \multicolumn{6}{l}{$^b$ Ensemble on 6 folds for each projection 18 models total}\\
          \multicolumn{6}{l}{$^c$ Ensemble of 6 models selected from 16 by ensemble result on test set.}\\
          \multicolumn{6}{l}{$^d$ 3 ensembles merged with coefficients selected on test set.}\\
        \end{tabular}
    \end{table*}
   
    \subsection{COVID-19 segmentation Models} 
    We use data described in Section \ref{sec:data} to train segmentation models that are able to localize COVID-19 lesions in lungs. We implemented the U-net with DPN-92 \citep{dpn92} and ResNet-21 \citep{he_resnet} as encoders, FPN with EfficientNet encoder \citep{EfficientNet} and a standalone ResNet-18 encoder. We, first, describe each of the networks in Sections 4.1.1 - 4.1.4 and then explain how we combined them into ensembles in Section \ref{ensembling}. 
 
        \subsubsection{DPN-92 U-Net.}
        We followed \citep{manvel2019radiologist} and trained the U-Net with a Dual Path Network (DPN) with 92 layers as an encoder with a lightweight decoder. Furthermore, we used the same learning rate, loss function, optimizer and augmentations,  as described in \citep{manvel2019radiologist}, only having the number of training steps reduced from 20,000 to 2,500.

        \subsubsection{Resnet-21 U-Net.}
        We trained the U-Net with ResNet-21 as an encoder \citep{he_resnet}. We used the Adam optimizer with the initial learning rate \num{3e-5} for the first 200 epochs and \num{1e-5} until convergence. The weight decay was \num{1e-4} for the whole training procedure. We used the batch size of 64 and the Dice measure as the loss function for this model \citep{v_net_dice_loss}. 
        
        \subsubsection{FPN with EfficientNet encoder}
        We also trained the Feature Pyramid Network model \citep{FPN} with the EfficientNet-B0 \citep{EfficientNet} encoder from the open source repository \citep{ptch_seg_models}. We used the Adam optimizer with the flat learning rate \num{3e-3} until convergence. The weight decay was \num{1e-8} for the whole training procedure. We used the batch size of 12 and binary cross-entropy as the loss function. 

        \subsubsection{Standalone ResNet-18}
        We also trained the ResNet-18 as a standalone segmentation model by removing the pooling and the fully connected layers. 
        We did it to diversify our ensemble by adding a different segmentation approach and examine whether plain convolution architecture like ResNet is able to extract features to handle the segmentation task.
        For the ResNet-18, we used the same hyperparameters and the training regime as for ResNet-21, except the batch size was reduced to 8. 
        
        \subsubsection{Preprocessing}
        Raw images from DICOM files are stored as 16-bit grayscale images. In order to make learning process more stable and robust we normalized input to the neural networks.
        For the DPN-92 U-Net and the FPN models we (see algorithm \ref{alg:normalize}):
        \begin{itemize}
            \item Multiplied the value of each pixel in the DICOM image array by the rescale slope and added the rescale intercept, these two parameters are stored in DICOM file format;
            \item Divided the result by the absolute value of the minimum pixel value in the image
            \item Truncated the value to the range [-0.505, 0.505].
        \end{itemize}
        
        \begin{algorithm}
         \SetKwInOut{Input}{input}\SetKwInOut{Output}{output}
         \Input{Raw DICOM pixel array $x$}
         \Output{Normalized pixel array $x_{\text{norm}}$ for the DPN-92 and FPN}
         \BlankLine
         $b$ -- rescale intersept, $k$ -- rescale slope, $p_{\text{min}} = |\min(x)|$\;
         \For{$\forall p \in x$}{
            $p \leftarrow \frac{p-b}{k}$\;
            $p \leftarrow p*\frac{1}{p_\text{min}}$\;
            \eIf{$p<-0.505$}{$p \leftarrow -0.505$}
            {\If{$p>0.505$}{$p \leftarrow 0.505$}
            }
         }
         \caption{Normalize input for DPN-92 and FPN networks}
         \label{alg:normalize}
        \end{algorithm}
        
        The data for the ResNet models was normalized to have zero mean and the unit standard deviation, i.e. we performed the following transformation for any input image $x$ from the training, validation and test\footnote{we emphasize that test dataset was collected from different medical centers} dataset:
        $$x \leftarrow (x - \mu)/\sigma,$$ where the mean $\mu$ and the standard deviation $\sigma$ for the normalization process were calculated on the training data from raw DICOM images.
        
        \subsubsection{Ensembling}\label{ensembling} 
        
        To improve on the quality of the individual models, we experimented with various ensembling techniques for each model as well as across the models.
        
        \begin{figure}
              \includegraphics[width=.45\textwidth]{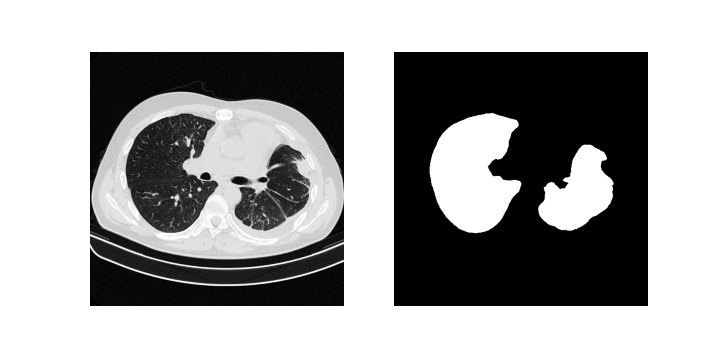}
              \includegraphics[width=.45\textwidth]{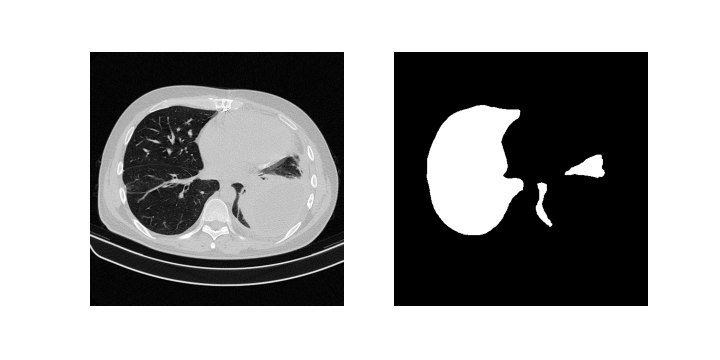}
            \caption{Examples of the lung segmentation (section \ref{subsec:segm_of_lungs})}
            \label{fig:lung_segm}
        \end{figure}
        
        For the DPN and FPN architectures, we trained 6 models each, selecting a different validation set form the training set for each model. The models were ensembled by averaging their predictions for each architecture.
        
        We trained 16 models for the Resnet-21 U-net and 2 models for the standalone ResNet-18 using random subsets of the training data. 
        For the ensemble model we take five of 16 ResNet-21 and the better of the two ResNet-18 models. We chose the best performing tuple of ResNet-21 over one thousand of $\binom{16}{5} = 4368$ randomly generated choices.
        The models were ensembled by the unanimous vote of all the models in the positive class.
        
        We also trained 12 additional DPN models in the sagittal and dorsal projections, an ensemble of 6 models for each projection, as described in \citep{manvel2019radiologist} and calculated performance of the 3-dimensional DPN ensemble.

\begin{figure*}
        \centering
          \includegraphics[width=\textwidth]{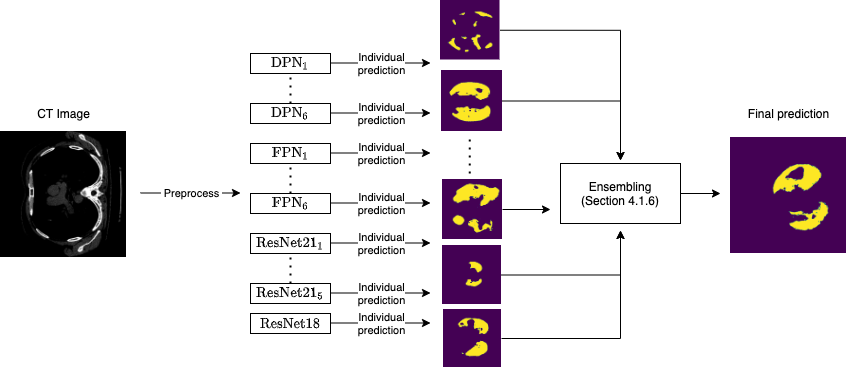}
          \caption{Inference for the individual slice.}
          \label{pipeline_fig}
        \end{figure*}

        To build the final ensemble used in the experiments, we modified the predictions of the best performing ensemble (5 Resnet-21 U-Net and one Resnet-18) with high confidence predictions from DPN-92 and FPN ensembles. The confidence thresholds for the final ensemble where tuned on the test set.
        Furthermore, we used the following scoring function for each pixel:
          
        +1 for each:
        \begin{itemize}
            \item predictions of all models in ResNet ensemble $>$ 0.5
            \item mean of DPN-92 U-Net models $>$ 0.7
            \item mean of FPN models $>$ 0.85
         \end{itemize}
        -1 for each:
        \begin{itemize}
            \item mean of DPN-92 U-Net models $<$ 0.3
            \item mean of FPN models $<$ 0.15
        \end{itemize}
        Pixels with positive values where considered as positive class predictions. The whole pipeline of the inference is depicted on figure ~\ref{pipeline_fig}.
        
        \subsection{Segmentation of Lungs}
        \label{subsec:segm_of_lungs}
        For the left and the right lung segmentation, we used FPN \citep{FPN} with the lightweight encoder EfficientNet-B0 \citep{EfficientNet}. The final segmentation was the result of ensemble of three separate 2-D networks for axial, coronal and sagittal projections (see figure \ref{fig:lung_segm}). In order to give the model better spatial understanding, we added 3D coordinates to the input as separate channels. We have also resized the inputs for all the projections to 128x128 pixels.
        
        The lung segmentation dataset contains numerous mistakes. To deal with them, we used active learning as follows. First, we trained the ensemble on the 50\% of the data and evaluated the results on the other 50\%. Second, the CT scans with the least dice scores were manually reviewed and masks errors were excluded from the dataset. Than, we repeated this process with the other half of the dataset. Overall, we excluded 19 CT scans from the dataset as the result of this cleaning process. After this, we trained the final networks with the holdout 10\% validation scheme. The resulting Intersection-over-Union (IoU) score for validation was 0.97, which is comparable to the human labeling quality of IoU = 0.96 \citep{segmentation-source}.

\section{Experiments}
\label{sec:experiments}
To evaluate the efficiency of the proposed models, we conducted a study based on the retrospective data collected during the treatment process in a large clinic in Russia. 
In this study, 58 chest CT scans on 49 patients were selected (see Section \ref{sec:data}) and were divided across the following four experiments described in the rest of this section. 

    \begin{table*}
    \centering
    
    \caption{Mean DSC and standard deviation on the segmentation experiment.}
    \label{exp_st_1_table}
            \centering
            \begin{tabular}{|p{3cm}|p{3cm}|p{3cm}|p{3cm}|p{2cm}|}
            \hline
            & Radiologist & Model & Number of cases & p-value  \\
            \hline
            Radiologist 1& 0.650(0.225) & \textbf{0.676(0.239)} & 20 & 0.728\\
            \hline
            Radiologist 2& 0.681(0.218) & \textbf{0.682(0.239)} & 20 & 0.985\\
            \hline
            Radiologist 3& 0.697(0.191) & \textbf{0.709(0.178)} & 20 & 0.832\\
            \hline
            Radiologist 4& 0.662(0.224) & \textbf{0.713(0.176)} & 20 & 0.443\\
            \hline
            Radiologist 5& \textbf{0.699(0.202)} & 0.686(0.235) & 20 & 0.855\\
            \hline
            Radiologist 6& 0.251(0.090) & \textbf{0.695(0.186)} & 20 & 0.000\\
            \hline
            All radiologists& 0.606(0.254) & \textbf{0.694(0.211)} & 120 & 0.004\\
            \hline
            \end{tabular}
        \end{table*}

    \subsection{Segmentation} 

    The goal of the first experiment was to compare the segmentation accuracy of pulmonary consolidation and the ground glass opacity area on the CT images obtained by our segmentation model vis-a-vis the performance of experienced radiologists involved in our study. For this purpose, we used the experimental subset F(20), which was represented by 20 CT cases from 20 patients of varying severity that was described in Section 4 and presented in Table~\ref{exp_dataset}.
    These cases were manually segmented by six experienced practicing radiologists and by our model.

    To measure the performance of an individual radiologist, we compared his or her results to the panel of remaining 5 radiologists. Each pixel was considered to belong to the positive class if at least 3 radiologists marked it as positive. We used the mean Dice similarity coefficient (DSC) for all the CT scans as our measurement metric. Since the panel result is different for each radiologist, we calculated the metric for our segmentation model separately for each panel.   

    As Table~\ref{exp_st_1_table} shows, our model outperforms 5 out of 6 radiologists and outperforms the average radiologist (represented by its last row) with statistical significance (p-value of 0.004).    
        
    \subsection{Patient's dynamics}
    
    In the second experiment, we compared performance of the segmentation model and human performance in assessing patient's dynamics for the follow-up CT-scans. For this purpose, we used the experimental subset S(18) (see Table \ref{exp_dataset}) consisting of 18 CT scans on 9 patients (2 CT scans per patient with different dates).
    The radiologists and our model independently estimated the percentage of lesions in the left and the right lung. Based on this information, one of the three classes for evaluating the patient dynamics was chosen by the radiologists and our model: a positive response to the therapy, disease progression and a stable condition (for our model the change of less than 1\% was considered to be stable).
    In case of one patient, the radiologists' assessments where tied 3 vs. 3 between the positive response and the stable condition. Therefore, we removed this case from the experiment since we were unable to determine the "ground truth" of this patient's dynamics.
    From the remaining 8 cases, the radiologists unanimously agreed on the dynamics of the disease progression in 7 cases, and in the remaining case they were split 5 to 1 in favor of the disease progression vs. the positive response. It turned out that our system correctly predicted the dynamics \emph{in all the 8 cases}.
    
    \subsection{Lesion share estimation}
    
        \begin{table*}
    \centering
    \caption{Lesion share estimation results of the radiologists and our model}
    \label{lesion-share}
            \centering
            \begin{tabular}{|p{2.5cm}|p{1cm}|p{1cm}|p{1cm}|p{1cm}|p{1cm}|}
            \hline
            & \multicolumn{2}{|l|}{Radiologist} & \multicolumn{2}{|l|}{Model} & \\
            \hline
            & MAE & ME & MAE & ME &Cases  \\
            \hline
            Radiologist 1& 0.08 & -0.06 & 0.13& -0.13 & 76\\
            \hline
            Radiologist 2& 0.10 & 0.10 & 0.11& -0.11 & 76\\
            \hline
            Radiologist 3& 0.06 & -0.03 & 0.13& -0.13 & 76\\
            \hline
            Radiologist 4& 0.07 & 0.03 & 0.12& -0.12 & 76\\
            \hline
            Radiologist 5& 0.09 & 0.07 & 0.11& -0.11 & 76\\
            \hline
            Radiologist 6& 0.12 & -0.11& 0.14  & -0.14 & 76\\
            \hline
            All radiologists& 0.09 & 0.00 & 0.13& -0.12 & 456\\
            \hline
            \end{tabular}
        \end{table*}

    In the third experiment we compared the performance of the segmentation model and human performance in assessing patient's lesion share which is the base for identification of the lung damage stage. 
    For that propose we used radiologists estimation of the lesions percentage in the left and the right lung made with the 5\% increment. In total, we had 76 estimations from six radiologists for the right and the left lung for each of the 38 chest CT scans from experimental subsets F(18) and T(20). We performed the same estimation for lesion share using our segmentation models to calculate the COVID-19 lesion and lungs volumes and dividing them. To measure the performance of an individual radiologist in comparison to our system, we compared their results to the panel of remaining 5 radiologists. As the ground truth, we took an average lesion share between the five remaining radiologists. We used the mean absolute error (MAE) for all the CT scans as our measurement metric. We also calculated the mean error (ME) to explore where there is a systemic component to the error.
    
\begin{figure}
\includegraphics[width=10cm]{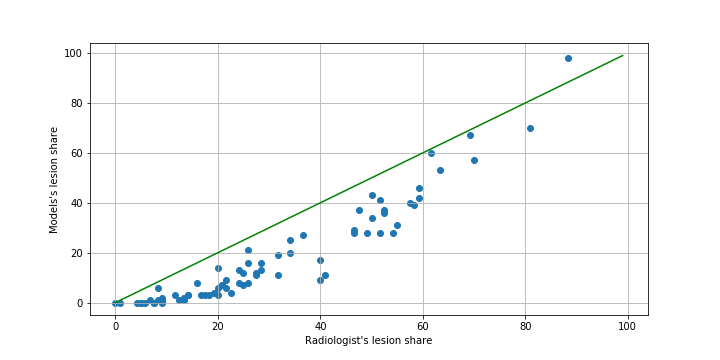}
\caption{Model's lesion share to radiologist's lesion share} \label{fig1}
\end{figure}

\begin{figure}
\includegraphics[width=8cm]{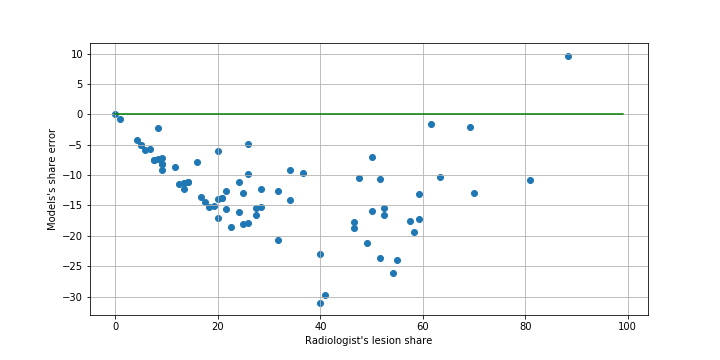}
\caption{Model's error to radiologist's lesion share} \label{fig2}
\end{figure}

\begin{figure}
\includegraphics[width=8cm]{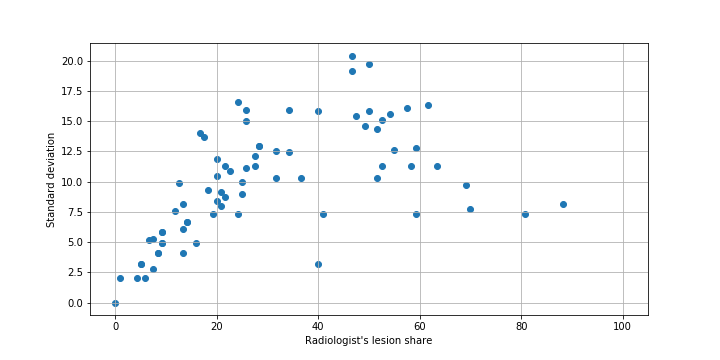}
\caption{Radiologist's lesion share standard deviation} \label{fig3}
\end{figure}

    As Table \ref{lesion-share} shows, that the radiologists estimation is significantly subjective, with some radiologists biased to overestimation (Radiologist 2) or underestimation (Radiologist 6). Our model estimates the lesion share based on objective factors and is highly correlated with the mean estimation of 6 radiologist per each of 76 considered cases (Fig. \ref{fig1}) while being biased to underestimation.
    
    The subjectivity of the radiologists estimate is correlated with lesion share. Lesion share range 30-70 has the largest disagreement in estimation as between the radiologists themselves (Fig. \ref{fig3}) as between the model and mean radiologist estimation (Fig. \ref{fig2}).
    
    Based on this, we conclude that the doctors make systemic biases in their estimations of lesion volumes while using typical diagnostic tools. As our analysis shows, our system corrects these estimation  biases and therefor is well-suited for estimating lesion volumes.

    \subsection{Classification}

    In the fourth experiment we compared the performance of the radiologists and our segmentation model results on the CT-classification task. We used classification accuracy as the metric for this experiment. For the classification task we used the segmentation model results. To estimate the CT-class, our system calculated the maximum share of the lesions in the right or the left lung and then used precalculated thresholds to get the final classification result.
    
    To correct for the radiologists subjective bias, we fitted the thresholds for lesion share for each CT class to maximise prediction accuracy. We split the experimental dataset into two equal folds stratified by the CT-classes retrieved form the hospital reports. For each radiologist we fitted the thresholds for the whole dataset and for each of the folds separately. 
    As can be seen from Table \ref{radiologist-bias}, the thresholds vary greatly between radiologists even when fitted on all the data: 0.03-0.12 for CT1-CT2, 0.24-0.44 for CT2-CT3 and 0.59-0.99 for CT3-CT4. When fitting on separate folds the individual thresholds become even more noisy due to lower number of data points avaivable.
    
    \begin{table*}
    \centering
    \caption{CT-class thresholds for radiologists}
    \label{radiologist-bias}
            \centering
            \begin{tabular}{|p{3cm}|p{1cm}|p{1cm}|p{1cm}|p{1cm}|p{1cm}|p{1cm}|p{1cm}|p{1cm}|p{1cm}|}
            \hline
            & \multicolumn{3}{|l|}{Split 1} & \multicolumn{3}{|l|}{Split 2} & \multicolumn{3}{|l|}{Full}\\
            \hline
             & CT-2 & CT-3 & CT-4 & CT-2 & CT-3 & CT-4 & CT-2 & CT-3 & CT-4  \\
            \hline
            Radiologist 1& 0.12 & 0.72 & 0.74 & 0.17 & 0.28 & 0.84 & 0.12 & 0.44 & 0.83\\
            \hline
            Radiologist 2& 0.03 & 0.25 & 0.47 & 0.03 & 0.22 & 0.84 & 0.03 & 0.24 & 0.83\\
            \hline
            Radiologist 3& 0.12 & 0.42 & 0.74 & 0.17 & 0.34 & 0.84 & 0.12 & 0.36 & 0.83\\
            \hline
            Radiologist 4& 0.03 & 0.25 & 0.74 & 0.06 & 0.30 & 0.59 & 0.06 & 0.24 & 0.59\\
            \hline
            Radiologist 5& 0.10 & 0.29 & 0.74 & 0.04 & 0.28 & 0.77 & 0.10 & 0.28 & 0.79\\
            \hline
            Radiologist 6& 0.10 & 0.25 & 0.99 & 0.26 & 0.55 & 0.84 & 0.10 & 0.29 & 0.99\\
            \hline
            All radiologists& 0.08 & 0.29 & 0.74 & 0.06 & 0.34 & 0.84 & 0.06 & 0.29 & 0.83\\
            \hline
            \end{tabular}
        \end{table*}

    To estimate the collective bias of the panel, we combine individual biases of each radiologist in the panel by fitting the optimal thresholds on all their estimations in the train split.
    As can be seen from Table \ref{CT-class_cor}, after applying correction for the panel bias, our system outperforms all the radiologists in the study, 3 of them with statistical significance (p-values of 0.01 or less).
    
\begin{table*}
    \centering

    \caption{CT-classification accuracy of the radiologists and our system}
    \label{CT-class_cor}
            \centering
            \begin{tabular}{|p{3cm}|p{15mm}|p{15mm}|p{15mm}|p{15mm} |p{15mm}|p{15mm}|p{15mm}|}
            \hline
            & \multicolumn{2}{|l|}{Split 1 (29 cases)} & \multicolumn{2}{|l|}{Split 2 (29 cases)} & \multicolumn{3}{|l|}{Combined (58 cases)} \\
           \hline
            & Radiologist & Model & Radiologist & Model & Radiologist & Model & p-value  \\
             \hline
            Radiologist 1& 0.76 & 0.86 & 0.76 & 0.83 & 0.76 & 0.84 & 0.248\\
            \hline
            Radiologist 2& 0.59 & 0.90 & 0.66 & 0.86 & 0.62 & 0.88 & 0.001\\
            \hline
            Radiologist 3& 0.72 & 0.90 & 0.83 & 0.90 & 0.78 & 0.90 & 0.080\\
            \hline
            Radiologist 4& 0.66 & 0.90 & 0.76 & 0.90 & 0.71 & 0.90 & 0.010\\
            \hline
            Radiologist 5& 0.66 & 0.83 & 0.86 & 0.72 & 0.76 & 0.78 & 0.828\\
            \hline
            Radiologist 6& 0.55 & 0.83  & 0.83 & 0.97 & 0.69 & 0.90 & 0.006\\
            \hline
            All radiologists& 0.66 & 0.87 & 0.78& 0.86 & 0.72 & 0.86 & \num{1,66E-06}\\
            \hline
            \end{tabular}
        \end{table*}

\subsection{Baseline Comparison}

We have also compared the performance of our model with the existing COVID-19 detection baselines. It turned out that many existing models, such as (\citet{weakly_supervised_framework, cell_paper, light_network, light_network2}), were incomparable with our approach both for the segmentation and the classification cases for the following reasons. 
First, some of the existing approaches did not provide code or data, while others used the classification and segmentation criteria that are different from our method, thus rendering them incomparable. For example, \citep{weakly_supervised_framework} used the binary (presence/absence of COVID-19) classification, whereas we deployed the CT-0/CT-1/CT-2/CT-3/CT-4 classification commonly used in Russia and some other countries. 
Similarly, we could not compare our approach to many segmentation methods due to the lack of the code and data.

The segmentation baselines comparable with our approach are the Inf-Net model described in \citep{infnet}, and the MiniSeg model described in \citep{light_network2}. We compare our model with these baselines in the rest of this subsection using two testing datasets. The first dataset is COVID-19 CT Segmentation Dataset \footnote{Available at \url{http://medicalsegmentation.com/covid19/}} \citep{jenssen2020covid} refered hereafter as CovidCTSegmentation, this dataset is very small containing only 100 CT slices from 60 patients. The second dataset is the MosMedData dataset mentioned earlier.

We compared the performance of MiniSeg model available on GitHub \footnote{Available at \url{https://github.com/yun-liu/MiniSeg}} and CoRSAI on CovidCTSegmentation training both models from scratch and using  5-fold cross-validation. The resulting DSC score was 0.452 for MiniSeg and 0.744 for CoRSAI showing excellent performance of our model even when presented with minimal amount of training data. 

\begin{figure*}
\includegraphics[width=\textwidth]{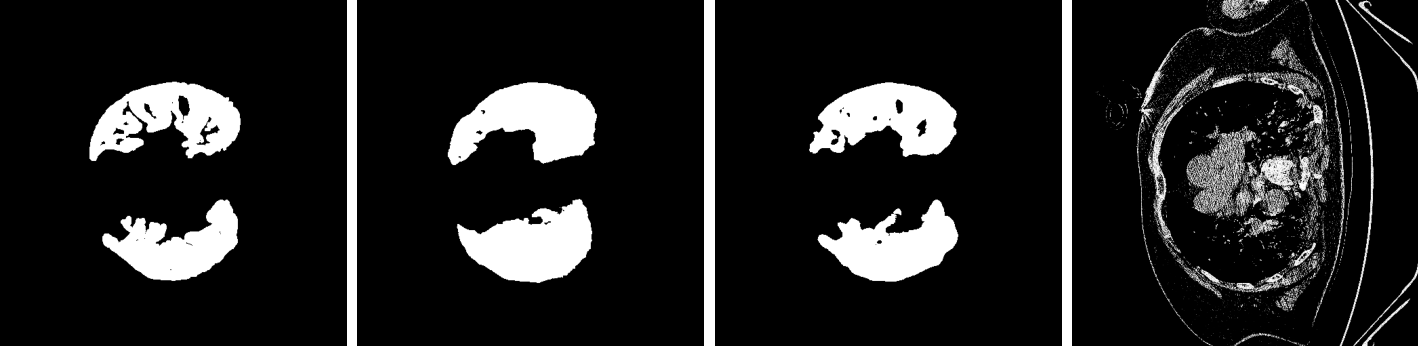}
\newline
\includegraphics[width=\textwidth]{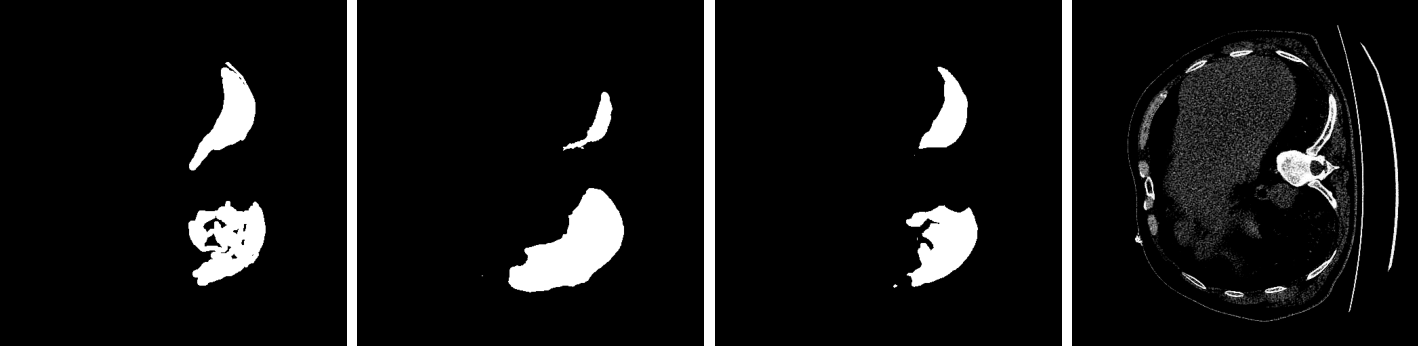}
\caption{Examples of the segmentation of a slice from CovidCTSegmentation dataset.
\newline
The leftmost image is the output of the MiniSeg model, next to the right is the output of the CoRSAI, next to the right is the ground truth, the rightmost image is the input slice}
\label{fig:segmentation_example}
\end{figure*}

We also examined the following cases comparing CoRSAI, MiniSeg and  various baselines on the CovidCTSegmentation dataset using 5-fold cross-validation as in \citep{light_network2}:
\begin{itemize}
    \setlength\itemsep{0em}
    \item Baseline results for (U-Net, Inf-Net, EfficientNet) as stated in \citep{light_network2}
    \item MiniSeg
    \item CoRSAI 
\end{itemize}

The results  are shown in Table  \ref{small-dataset-comparison}, where the baseline DSC performance scores are taken directly from (Qiu et al. 2020). As Table \ref{small-dataset-comparison} shows, the CoRSAI model outperformed the baselines in terms of the DSC metric.

\begin{table}[!t]
\centering
\caption{Comparison on CovidCTSegmentation dataset}
\label{small-dataset-comparison}
            \begin{tabular}{|p{5cm}|p{2cm}|}
            \hline
            \textbf{Model}&\textbf{DSC score}\\ 
            \hline
            U-Net & 0.684\textsuperscript{*}\\
            Inf-Net & 0.744\textsuperscript{*}\\
            EfficientNet & 0.705\textsuperscript{*}\\
            MiniSeg & 0.759\textsuperscript{*}\\
            \hline
            CoRSAI & \textbf{0.768}\\
            \hline
            \multicolumn{2}{l}{\textsuperscript{*}\footnotesize{Results from \citep{light_network2}}}
            \end{tabular}

        \end{table}
        
Next we compared the Inf-Net, MiniSeg and CoRSAI models on MosMedData dataset. We took the initial Inf-Net model, as available on GitHub, including its architecture and the computed weights, and tested it vis-a-vis our model on the MosMedData dataset in terms of the Dice performance metric. It turned out that the "as-is" Inf-Net model produced only 0.195 Dice metric on MosMedData, which was significantly below our model having the value of 0.643 for the Dice metric. 
This inferior performance of Inf-Net was due to the fact that Inf-Net was trained on the very different dataset obtained for the Wuhan COVID-19 patients.        
        
To provide further comparison of the three models, we compared CoRSAI, MiniSeg and Inf-Net on the MosMedData dataset using our private training dataset to retrain and fine-tune all the models using the same methods, as we did for our model described in Section \ref{sec:methods}.
The results  are shown in Table  \ref{mosmeddata-dataset-comparison} from which it is clear that CoRSAI outperformed Inf-Net and MiniSeg on the MosMedData dataset.

\begin{table}[!t]
\centering
\caption{Comparison on MosMedData dataset}
\label{mosmeddata-dataset-comparison}
            \begin{tabular}{|p{5cm}|p{2cm}|}
            \hline
            \textbf{Model}&\textbf{DSC score}\\
            \hline
            MiniSeg & 0.032\\
            Inf-Net & 0.619\\
            CoRSAI & \textbf{0.643}\\
            \hline
            \end{tabular}

        \end{table}

As a result of this extensive retraining, we improved the performance of Inf-Net from 0.195 to 0.619, which is in line with the performance results of the joint ResNet-21+ResNet-18 and DPN-3D ensembles  presented in Table \ref{lesion-share}, such as DPN and FPN, but still inferior to the 0.643 of our model.
The MiniSeg model achieved DSC score of 0.032 after the same retraining. 
We need to mention that for the MiniSeg retraining we used all hyperparameters ``as-is'' excluding the batch size. In our retraining we set the batch size to 24 to reduce time needed for the training. Dataset that we used to retrain MiniSeg model is much bigger than the original dataset used in \citep{light_network2} used for the training MiniSeg. Since MiniSeg is extremely small network with approximately 86000 parameters this might be a problem of catastrophic forgetting \citep{catastrophic_forgetting} and hence a reason for generalization performance on our data.

We maintain that the superior performance of our model is due to  the careful deployment of ensembles for both the individual models (DPN, FPN, ResNet-21, etc.) and for combining of these individual models into one ensemble, as described in Section \ref{ensembling} and shown in Table \ref{lesion-share}.

\section{Conclusion}
\label{sec:conclusion}
In this paper we described a novel ensemble of previously proposed deep convolutional neural networks specifically modified for the COVID-19 induced pneumonia segmentation task for the analysis of the chest CT scans and the corresponding CoRSAI system. Furthermore, we have created a segmentation-based classification model to categorize the severity level of the disease on those CT scans. 
To test the performance of our models, we conducted an experiment that showed that our model outperformed most of the experienced radiologists in the segmentation and all the radiologists in the classification tasks. It also managed to predict the dynamics of the disease with the 100\% accuracy.

Our model has been favorably received by the medical community in Russia and has been recently deployed in several hospitals in the country. 

In particular, CoRSAI is publicly available for the doctors and anybody else who is interested in our system on the website \url{https://ai.sberhealth.ru/covid19/}. Moreover, it was deployed in 46 medical institutions in 25 different regions of the Russian Federation and thousands of the CT studies have been processed with its help since May 2020. Although some discrepancies were highlighted between radiologists and the model in lesion estimation, the system demonstrated good acceptance by the medical community. It was emphasized that the system has speeded up the lesion estimation time and improved its accuracy, which is particularly important for evaluating patients dynamics.
However, doctors have also noted certain limitations of our system, and fixing these limitations is a part of the future plans for working on CoRSAI, including problems of differentiation of the ground glass and consolidation, assessment of the type of pneumonia (bacterial or viral), as well as extending the list of lung pathologies for diagnosis (tuberculosis, emphysema, lung cancer, pneumathorax, etc).

Therefore, as a part of the future work, we plan to measure the performance of this deployed model in the actual clinical settings in terms of how much it helps the doctors to treat the coronavirus patients. We also plan to fine-tune and further improve the model based on this feedback.

\bibliographystyle{ACM-Reference-Format}
\bibliography{sample-base}


\end{document}